\documentclass{article}
\usepackage{epsf}
\usepackage{graphicx}
\usepackage{eepic}
\usepackage{color}
\usepackage[colorlinks]{hyperref}
\usepackage{latexsym,epsfig}
\def\sl{\llap{$/$}}
\newcommand{\be}{\begin{equation}}
\newcommand{\ee}{\end{equation}}
\newcommand{\ba}{\begin{array}{c}}
\newcommand{\ea}{\end{array}}
\newcommand{\bqa}{\begin{eqnarray}}
\newcommand{\eqa}{\end{eqnarray}}

\begin{document}

\begin{center}

{\Large\bf Investigations on the Property of $f_0(600)$ and
$f_0(980)$ Resonances in  $\gamma\gamma\to \pi\pi$
Process}\footnote{Talk given by H. Q. Zheng at 6th International Workshop on Chiral Dynamics, CD09\\
        July 6-10, 2009,
        Bern, Switzerland}
 \vspace{10mm}

{\sc X.~G.~Wang,$^1$ O.~Zhang,$^1$ L.~C.~Jin,$^1$ H.~Q.~Zheng,$^1$
Z.~Y.~Zhou$^2$}
\\[5mm]

\noindent {\it 1: Department of Physics, Peking University, Beijing
100871, China}
\\
\noindent {\it 2:Department of Physics, Southeast University,
Nanjing 211189, China }
\\[10mm]
\today
\end{center}

\abstract{Using dispersion relation technique and experimental data,
a coupled channel analysis on $\gamma\gamma\to\pi\pi$ process is
made. Di-photon coupling of $f_0(600)$ and $f_0(980)$ resonances are
extracted and their dynamical properties are discussed. Especially
we study the physical meaning of the coupling constant
$g^2_{\sigma\pi\pi}$, which maintains a negative real part as
determined through  dispersive analyses.}

\section{A dispersive analysis on $\gamma\gamma\to \pi^+\pi^-, \pi^0\pi^0$ processes }

In recent few years  there have been renewed interests on the study
of the $\gamma\gamma\to \pi\pi$ process, partly due to the new
experimental data provided  by Belle Collaboration.~\cite{Belle} The
investigation on such a process enables us to extract the di-photon
coupling of resonances appearing in this reaction, which, as
emphasized by Pennington,~\cite{Penn072} affords a unique
opportunity in exploring the underlying  structure of these states.
Along with previous work found in the
literature,~\cite{ggpipi,prades,ochs} we performed a dispersive
analysis on $\gamma\gamma\to \pi^+\pi^-, \pi^0\pi^0$
processes.~\cite{Mao08} The major differences between
Ref.~\cite{Mao08} and much of previous work is that in the former we
try to perform a coupled channel analysis in the strongly
interacting I=0 $s$-wave -- hence information from $\bar KK$ channel
is also taken into account, at least in principle. We also fit Belle
data up to 1.4GeV, which is certainly useful in fixing the
$d$-waves. A better determination to the $d$-waves turns out to be
very important in studying the low energy $s$-waves as well, where
$d$-waves serve as a background contribution.

The dispersion representation of $\gamma\gamma\to\pi\pi,\bar KK$
amplitudes, $F(s)$, takes the following form:~\cite{Basdevant}
\begin{equation}
F(s)=F_{B}+D(s)[Ps-\frac{s^{2}}{\pi}\int_{4m_{\pi}^{2}}
\frac{\mathrm{Im}D^{-1}(s')F_{B}(s')}{s'^{2}(s'-s-i\epsilon)}ds']\ ,
\end{equation}
where $F_B$ denotes the Born term, $P$ is a two dimensional
(subtraction) constant array. The $2\times 2$ matrix function $D(s)$
obeys the following equation:
\begin{eqnarray}\label{D}
 D(s)&&=D(0)+
 \frac{s}{\pi}\int_{4m_{\pi}^{2}}\frac{D(s')\rho(s')T^{*}(s')}{s'(s'-s-i\epsilon)}ds'\
 ,
 \end{eqnarray}
where $\rho=\mathrm{diag}(\rho_1,\rho_2)$ and
$\rho_1=\sqrt{1-4m_\pi^2/s}$, $\rho_2=\sqrt{1-4m_K^2/s}$,
respectively; $T(s)$ denotes the $2\times 2$ partial wave $\pi\pi,
\bar KK$ scattering amplitudes. Numerical solution of Eq.~(\ref{D})
can be searched for. In the degenerate case of single channel
problem, function $D$ in Eq.~(\ref{D}) has a well-known analytic
representation -- the Omn\'es solution:
 \begin{equation}
 D(s)=\exp\left(\frac{s}{\pi}\int_{4m_\pi^2}^{\infty}\frac{\delta(s')ds'}{(s'-s)s'}\right)\
 .
 \end{equation}

 The $s$-wave  $T$ matrix in Eq.~(\ref{D}) is obtained by fitting a
 coupled channel $K$ matrix~\cite{AMP} to
 data.~\cite{datapiK,pislak} The relevant poles are listed
 in table~\ref{tab1}.
\begin{table}[hbt]
\centering\vspace{0.1cm}
\begin{tabular}{|c|c|c|}
\hline pole& sheet--II & sheet--III
\\ \hline
$\sigma$ & $0.549-0.230i$   & -
\\ \hline $f_0(980)$& $0.999-0.021i$  & $0.977-0.060i$ i\\
\hline
\end{tabular}
\caption{\label{tab1}The pole locations on the $\sqrt{s}$--plane, in
units of GeV.
}
\end{table}
We notice from table~\ref{tab1} that the $f_0(980)$ resonance may
consist of two poles -- one locates on sheet II, while the other on
sheet III, though the latter is found  not quite stable in the
numerical fit. Though the twin-pole phenomenon with respect to
$f_0(980)$ was mentioned long time ago,~\cite{kato} in
$\gamma\gamma\to\pi\pi$ process one discovers further evidence in
support of the idea that the $f_0(980)$ resonance could be  a
coupled channel Breit--Wigner resonance.~\cite{Mao08} Similar
phenomenon may occur in the situation of $X(3872)$
particle.~\cite{zhangou}

The two I=0 $d$-wave and the I=2 $s$-wave amplitudes are attained
through single channel approximation and the corresponding $\pi\pi$
scattering $T$ matrices are borrowed from
Refs.~\cite{pipiDwave,pku3}. With these $T$ matrices the Omne\'s
solution is used to determine the corresponding $D$ functions.
Other partial waves are tiny and have been approximated by their
Born terms. Then the $\gamma\gamma\to \pi^+\pi^-,\pi^0\pi^0$
cross-sections can  be fitted and the di-photon coupling of
$f_0(600)$, $f_0(980)$, $f_2(1270)$ resonances can be extracted. We
refer to Ref.~\cite{Mao08} for the numerical results and related
discussions.

By re-analyzing the whole process, the above estimates can be
advanced, especially $at$ $lower$ $energies$. An improved I=0
$s$-wave single channel $\pi\pi$ scattering $T$ matrix~\cite{pku3}
provides a better analyticity property than that of a usual $K$
matrix formalism, and gives a $\sigma$ pole location in nice
agreement with the Roy equation analysis.~\cite{CCL} The extracted
di-photon width $\Gamma(\sigma\to 2\gamma)\simeq 2.1$keV -- a number
significantly smaller than the value one expects for a naive $\bar
qq$ meson. Therefore the result indicates the non-$\bar qq$ nature
of the $f_0(600)$ resonance.

In the calculation as described by the last paragraph, as a
byproduct when extracting the di-photon coupling one also gets the
$\sigma\pi\pi$ coupling:
 \be\label{sigmapipisingle}
g^2_{\sigma\pi\pi}=(-0.20-0.13i)\mathrm{GeV}^2 \ .
 \ee
It could be surprising to notice that the real part of the coupling
strength, $\mathrm{Re}[g^2_{\sigma\pi\pi}]$, is negative. A narrow
resonance with such a property is not allowed, since it would be a
ghost rather than a particle. \footnote{The value, and especially
the sign given in Eq.~(\ref{sigmapipisingle}) is in qualitative
agreement with that of Ref.~\cite{ochs} and especially
Ref.~\cite{prades}. Notice that in Ref.~\cite{prades} there is a
sign difference in the definition of coupling strength. }
 In the next section
we devote to the discussion on physics behind this (once again) odd
property of the $f_0(600)$ or
 $\sigma$ meson.

\section{What does a negative \protect{$\mathrm{Re}[g^2_{\sigma\pi\pi}]$} tell us?}

\begin{figure}[h]
\centering
\includegraphics[height=6cm,width=12.5cm]{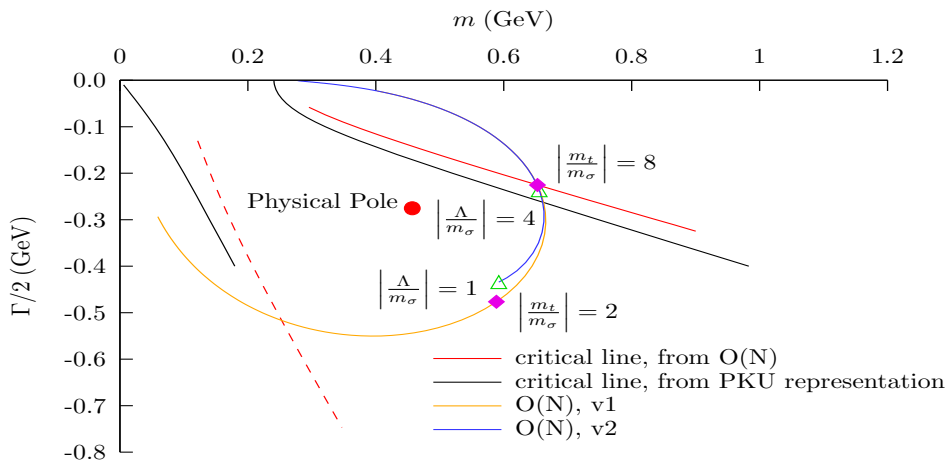}
\caption{\label{fig_couple}Region on the $\sqrt{s}$-plane with
$\mathrm{Re}[g^2_{\sigma\pi\pi}]<0$. }
\end{figure}

The negative value of  $\mathrm{Re}[g^2_{\sigma\pi\pi}]$ is related
to the large width of $f_0(600)$ meson. To initiate the
investigation let us recall the PKU dispersive representation for a
partial wave elastic scattering $S$ matrix element:~\cite{pku3,pku1}
 \be
 S^{phy.}=\prod_iS^{R_i}\cdot S^{cut}\ ,
\ee
 where $S^{R_i}$ denotes the $i$-th resonances on sheet II and $S^{cut}$ stands for the
 cut contribution. In each $S^{R_i}$ pole residue is a function of
 the pole location, and hence if we neglect every pole and cut
 contribution other than the $f_0(600)$ pole, we can obtain its
 coupling strength to two pions, $g^2_{\sigma\pi\pi}=({-0.18}-0.20i)\mathrm{GeV}^2$, which is found
 not much different from the value given by
Eq.~(\ref{sigmapipisingle}). This implies that the $\sigma\pi\pi$
coupling is mainly of a kinematical effect, $i.$ $e.$, largely
affected by the $\sigma$ pole location. In Fig.~\ref{fig_couple} we
draw the region where the residue contains a negative real part
based on the above approximation, $i.$ $e.$, considering only single
pole contribution. In the following, however, by studying the
solvable $O(N)$  $\sigma$ model, we will be able to learn more
lessons on physics of negative coupling strength.

The bare IJ=00 channel $\pi\pi$ scattering amplitude takes the
following form:~\cite{coleman74}
$$
T^{00}\left(s\right) =\frac{1}{32\pi}
\frac{s-m_\pi^2}{f_\pi^2-\left(s-m_\pi^2\right)\left({ \frac{1}
\lambda_0+\widetilde{B}_0\left(s\right)}\right)} \label{e:Tb}
$$
where
$$
\widetilde{B}_0\left(p^2\right)=\frac{-i}{2} \int\frac{\mathrm{d}^4
q}{\left(2\pi\right)^4}\frac{1}{q^2-m_\pi^2}\frac{1}{\left(p+q\right)^2-m_\pi^2}
$$
is a divergent integral and can be made finite by redefining the
renormalized coupling constant as,~\cite{chivu}
 \bqa\label{runningM}
\frac{1}{\lambda(M)}=\frac{1}{\lambda_0}-\frac{i}{2}\int\frac{d^4q}{(2\pi)^4}\,
\frac{1}{(q^2+i\epsilon)(q^2-M^2+i\epsilon)}\ , \eqa
$$\label{lambdaMFT} {
\frac{1}{\lambda(M)}+\tilde B(p^2;M)}=\frac{1}{\lambda_0}+\tilde
B_0(p^2) \ ,$$ where
 $$ \tilde B(s;
M)=\frac{1}{32\pi^2}\left[1+\rho(s)\log\frac{\rho(s)-1}{\rho(s)+1}-\log\frac{m_\pi^2}{M^2}\right]\
.
$$

To define the theory one can set
$${1\over\lambda(M)}=0\ ,$$
where $M$ denotes the scale when perturbation expansion fails,
though above the scale $M$ the theory can still be fine. The RGE of
coupling constant $\lambda$ becomes exact,
 \be\label{RGEON}
\mu^2\frac{d\lambda}{d\mu^2}=\frac{\lambda^2(\mu^2)}{32\pi^2}\ . \ee
The true problem of such a theory (herewith  called as $O(N)$ v1) is
that a tachyon appears at $m_t^2$, and hence the theory only works
when $|s|<<|m_t^2|$.~\cite{chivu}

If one does not like the tachyon  a sharp momentum cutoff at
$\Lambda$ can be used to make the theory finite. In this way one
avoids the tachyon, but a  spurious cut (at $4\Lambda^2$) and a
spurious physical sheet pole near the spurious cut occur, instead.
By this mean we define a  cutoff version of the effective theory.
Setting for example
$${1\over\lambda(\Lambda)}=0\ ,$$
defines   another version of $O(N)$ model (called as $O(N)$ v2
hereafter).

The region where $\mathrm{Re}[g^2_{\sigma\pi\pi}]<0$ is plotted in
Fig.~\ref{fig_couple} both for $O(N)$ model v1 and v2, which are,
however, almost identical. The $\sigma$ pole trajectories with
respect to varying the defining scale of two models are also
plotted. Clearly, seen from Fig.~\ref{fig_couple},  it is actually
very difficult for $O(N)$ models to reach the `realistic' $\sigma$
pole location. In model v1, one has to decrease the scale $M$ to
face a situation that the tachyon pole mass and the $\sigma$ pole
mass are comparable in magnitude, and hence breaks down the validity
of the effective theory. In model v2 similar things happen; in order
to get the $\sigma$ pole deep inside the region where
$\mathrm{Re}[g^2_{\sigma\pi\pi}]<0$, one has to decrease the cutoff
parameter  $\Lambda$ facing the situation that the $\sigma$ mass
 is comparable in magnitude with $\Lambda$, and thus  also
results in breaking the validity of the effective theory. The
conclusion is that QCD interaction in the scalar sector becomes so
strong that,
 the $O(N)$ toy model even fails
to handel the situation when the $\sigma$ pole gets as light and
broad as it is determined from reality. A more `realistic'
calculation also leads to a similar conclusion.~\cite{xiao05}

Another way to look at the non-perturbative nature of the $\sigma$
meson is through examining  the renormalization group equation,
Eq.~(\ref{RGEON}). To get the `realistic' $\sigma$ pole location,
one finds $\lambda(\mu)$ blows up at $\mu\simeq 0.55$MeV.

It is certainly an extremely  hard and non-perturbative task to
predict a pole from an effective lagrangian inside which the pole
does not have a corresponding field. Such kind of poles are
sometimes called as `dynamically generated' resonances. Once the
existence of the $\sigma$ pole was firmly established, it is
wondered  whether one should add the $\sigma$ field explicitly into
the low energy effective lagrangian. However the blow up of the the
coupling constant $\lambda$ at very low energy indicates that, even
if  the explicit $\sigma$ degrees of freedom is added into the
effective lagrangian, one still face a strongly non-perturbative
problem.

To summarize,  the $\sigma$ pole manifests the maximal
`non-perturbativity' that QCD could offer.

This work is supported in part by National Nature Science Foundation
of China under Contract
Nos. 10875001, 
10721063, 
10647113 and 10705009.


\begin{thebibliography}{99}
\bibitem{Belle}T.~Mori \textit{et al.} (Belle Collaboration), Phys. Rev.
{\bf D75}(2007)051101.
\bibitem{Penn072}
 M.~R.~Pennington, invited talk at YKIS Seminar on \textit{New Frontiers in QCD: Exotic Hadrons and Hadronic Matter},
 Kyoto, Japan, 20 Nov - 8 Dec 2006.
 Prog. Theor. Phys. Suppl. {\bf 168}(2007)143.
 \bibitem{ggpipi}Here we are only able to provide an incomplete list
 of references:
 D.~Morgan, M.~R.~Pennington, Z.~Phys. {\bf
 C48}(1990)623;
G.~Mennessier, Z.~Phys. {\bf C16} (1983) 241; A.~V.~Anisovich,
V.~V.~Anisovich, Phys. Lett. {\bf B467} (1999) 289; L.~V.~Fil'kov,
V.~L.~Kashevarov, Phys. Rev. {\bf C72} (2005) 035211; N.~N.~Achasov,
G.~N.~Shestakov, arXive: 0712.0885 [hep-ph]; J.~A.~Oller, L.~Roca,
C.~Schat, Phys. Lett. {\bf B659} (2008) 201.
\bibitem{prades}
J.~Bernabeu, J.~Prades, Phys. Rev. Lett. {\bf 100} (2008) 241804.
\bibitem{ochs}
G.~Mennessier, S.~Narison, W.~Ochs, Phys. Lett. {\bf B665} (2008)
205.

 \bibitem{Mao08}Y.~Mao et al., Phys. Rev. {\bf D79} (2009) 116008.
 \bibitem{Basdevant}O.~Babelon et al.,  Nucl. Phys.
{\bf B113}(1976)445; O.~Babelon et al.,  Nucl. Phys. {\bf
B114}(1976)252.
\bibitem{AMP}K.~L.~Au, D.~Morgan and M.~R.~Pennington, Phys. Rev.
{\bf D35}(1987)1633.
\bibitem{datapiK}W.~Ochs, Ph.D. thesis, Munich Univ., 1974;
D.~H.~Cohen  \textit{et al}.,  Phys. Rev. {\bf D22}(1980)2595;
A.~Etkin  \textit{et al}.,  Phys. Rev. {\bf D25}(1982)1786; A. D.
Martin, E. N. Ozmutlu,  Nucl. Phys. B158(1979)520;  G.~Costa
\textit{et al}.,  Nucl. Phys. {\bf B175}(1980)402;
V.~A.~Polychronakos  \textit{et al}.,  Phys.~Rev. {\bf
D19}(1979)1317.
\bibitem{pislak} S.~Pislak \textit{et al}.,
Phys. Rev. {\bf D67}(2003)072004.
\bibitem{kato}
Y.~Fujii, M.~Kato, Nuovo Cimento {\bf 13A}(1973)311; D.~Morgan,
Nucl.~Phys.\ {\bf A543}(1992)632.
\bibitem{pipiDwave}J.~J.~Wang, Z.~Y.~Zhou, H.~Q.~Zheng, JHEP {\bf 0512}(2005)019.
\bibitem{zhangou}O. Zhang et al.,
arXiv: 0901.1553 [hep-ph], to appear in PLB.
\bibitem{pku3}Z.~Y.~Zhou \textit{et al}., JHEP 0502(2005)043.
\bibitem{CCL}I.~Caprini,
G.~Colangelo, H.~Leutwyler,   Phys. Rev. Lett. 96, 132001 (2006).
\bibitem{pku1}
 Z.~Y.~Zhou, H.~Q.~Zheng, Nucl.
Phys.{\bf A775}(2006)212;  H.~Q.~Zheng et al., Nucl. Phys. {\bf
A733}(2004)235;
\bibitem{coleman74}S.~Coleman, R.~Jackiw and H.~D.~Politzer, Phys.
Rev. {\bf D10}(1974)2491.
\bibitem{chivu}R.~Sekhar~Chivkula and M.~Golden, Nucl. Phys. {\bf
B372}(1992)44 and references therein.
\bibitem{xiao05}
M.~X.~Su, L.~Y. Xiao, H.~Q.~Zheng, Nucl. Phys. {\bf A792}(2007)288.
\end{thebibliography}
\end{document}